\documentclass[aps,prd,a4paper,preprint,preprintnumbers,titlepage,
nofootinbib,superscriptaddress]{revtex4-2}

\usepackage[a4paper, hdivide={1.91cm,,1.165cm},
vdivide={1.83cm,,3.0cm}]{geometry}

\usepackage[normalem]{ulem}

\usepackage{graphicx}

\usepackage[hyperfootnotes=false,colorlinks=true,citecolor=blue,urlcolor=blue]{hyperref}

\usepackage{amsfonts}

\usepackage{amsmath}
\usepackage{amssymb}

\usepackage{bbm}

\usepackage{bbold}

\usepackage{adjustbox}
\usepackage{booktabs}

\usepackage{colortbl}

\usepackage{cancel}

\usepackage{color}
\usepackage{xcolor}

\usepackage{graphicx}
\usepackage{xspace}

\usepackage{units}

\usepackage{slashed}

\usepackage{tensor}

\usepackage{gensymb}

\usepackage{cleveref}

\usepackage{tabularx}
\usepackage{multirow}

\usepackage{setspace}
\usepackage{orcidlink}

\usepackage{caption}
\usepackage{subcaption}

\usepackage{enumerate}
\usepackage{enumitem}

\usepackage[version=3]{mhchem}

\usepackage{lipsum}

\newcommand{{\ia} }{{\i}}
\newcommand{{\Ia} }{{\.I}}

\def\s2tw{{\rm sin ^2 \theta_{W}}}

\def\beq{\begin{equation}}
\def\eeq{\end{equation}}
\def\bea{\begin{eqnarray}}
\def\eea{\end{eqnarray}}

\def\bea{\begin{eqnarray}}
\def\eea{\end{eqnarray}}
\def\beeq{\begin{eqnarray}}
\def\eeeq{\end{eqnarray}}

\def\ba{\begin{array}}
\def\ea{\end{array}}

\def\xis0{{\Xi^{*0}}}

\def\g5{\gamma_5}

\allowdisplaybreaks[1]
\newcommand{\Ds}{D_s}

\newcommand{\Dsone}{D_{s1}}
\newcommand{\Bs}{B_s}

\newcommand{\Bsone}{B_{s1}}
\newcommand{\GeV}{\mathrm{GeV}}
\newcommand{\MeV}{\mathrm{MeV}}
\newcommand{\keV}{\mathrm{keV}}
\newcommand{\ssbar}{\langle\bar s s\rangle}

\makeatletter
\def\frontmatter@abstract@produce{%
  \par
  \addvspace{\frontmatter@preabstractspace}%
  \begingroup
    \dimen@\baselineskip
    \setbox\z@\vtop{\unvcopy\absbox}%
    \advance\dimen@-\ht\z@\advance\dimen@-\prevdepth
    \@ifdim{\dimen@>\z@}{\vskip\dimen@}{}%
  \endgroup
  \begingroup
    \prep@absbox
    \unvbox\absbox
    \post@absbox
  \endgroup
  \@ifx{\@empty\mini@notes}{}{\mini@notes\par}%
  \addvspace\frontmatter@postabstractspace
}%
\makeatother

\begin{document}

\title{Mixing effects in radiative decays of heavy-strange axial-vector mesons within light-cone QCD sum rules}

\author{T.~M.~Aliev\,\orcidlink{0000-0001-8400-7370}}
\email{taliev@metu.edu.tr}
\affiliation{Department of Physics, Middle East Technical University,
Ankara, 06800, Turkey}

\author{S.~Bilmis\,\orcidlink{0000-0002-0830-8873}}
\email{sbilmis@metu.edu.tr}
\affiliation{Department of Physics, Middle East Technical University,
Ankara, 06800, Turkey}
\affiliation{TUBITAK ULAKBIM, Ankara, 06510, Turkey}

\author{M.~Savci\,\orcidlink{0000-0002-6221-4595}}
\email{savci@metu.edu.tr}
\affiliation{Department of Physics, Middle East Technical University,
Ankara, 06800, Turkey}

\date{\today}

\begin{abstract}
We calculate the radiative decay widths of
$\Dsone(2460)\to\Ds\gamma$, $\Dsone(2536)\to\Ds\gamma$,
$B_{s1}(5750)\to\Bs\gamma$, and $\Bsone(5830)\to\Bs\gamma$
within light-cone QCD sum rules, taking into account the mixing between the
$^{1}P_{1}$ and $^{3}P_{1}$ configurations. {In the bottom-strange
sector, we consider the experimentally established $\Bsone(5830)$ together
with the theoretically expected lower $1^+$ state, denoted
$B_{s1}(5750)$.} We obtain
\begin{align*}
\Gamma[\Dsone(2460)\to\Ds\gamma]
&=
{45.9_{-8.5}^{+10.1}}~\keV,
&
\Gamma[\Dsone(2536)\to\Ds\gamma]
&=
{0.55_{-0.48}^{+1.73}}~\keV,
\\
\Gamma[B_{s1}(5750)\to\Bs\gamma]
&=
{12.0_{-2.3}^{+2.9}}~\keV,
&
\Gamma[\Bsone(5830)\to\Bs\gamma]
&=
{8.9_{-2.0}^{+2.4}}~\keV.
\end{align*}
The $\Dsone(2460)\to\Ds\gamma$ decay width is substantially larger than that
of $\Dsone(2536)\to\Ds\gamma$. This difference originates from
$^{1}P_{1}$--$^{3}P_{1}$ mixing: the two basis-current contributions
interfere constructively for the lower state and destructively for the higher
state. {In the bottom-strange sector, the two widths are comparable
within their uncertainties, although the higher state still exhibits
destructive interference.} These
results show that the radiative decay pattern of strange heavy axial-vector
mesons {is sensitive to} the mixing of the two spin configurations.
\end{abstract}

\maketitle

\section{Introduction}
\label{sec:intro}

Radiative transitions of heavy-light mesons provide complementary information
on their internal structure. In particular, electromagnetic transitions are
sensitive to the spin composition of the states and to the coupling of the
photon to both the heavy and light quarks. They therefore complement mass
spectroscopy and strong-decay measurements in determining the properties of
heavy-light mesons.

The axial-vector ($J^P=1^+$) heavy-light mesons are particularly interesting
in this respect. Since charge conjugation is not a good quantum number for
heavy-light systems, the physical states are not pure $^{1}P_{1}$ or
$^{3}P_{1}$ configurations~\cite{Matsuki:2010zy}. Instead, spin-dependent interactions mix these
two basis states. In the heavy-quark limit, the physical states are classified
according to the total angular momentum of the light degrees of freedom,
$j_\ell=\frac{1}{2}$ and $j_\ell=\frac{3}{2}$, which correspond to definite
linear combinations of the $^{1}P_{1}$ and $^{3}P_{1}$ configurations.
For finite heavy-quark masses, deviations from this limit modify the mixing
pattern and consequently affect the radiative transition amplitudes. Since
both components contribute to the physical amplitudes, constructive or
destructive interference may occur, making the radiative decays sensitive to
the mixing between the two configurations.

The charm-strange axial-vector mesons $\Dsone(2460)$ and $\Dsone(2536)$ provide
a useful system for studying this effect. The $\Dsone(2460)$ is a very narrow
state lying below the $D^{\ast}K$ threshold, and its radiative decay to
$\Ds\gamma$ has been observed both in $B$ decays and in continuum production
\cite{BaBar:2004zjt,BaBar:2006eep}. The $\Dsone(2536)$, on the other hand, is
well established through hadronic decay modes such as $D^{\ast}K$ and
$D\pi K$ \cite{Belle:2007hue,BaBar:2011zpy}, while its radiative decay to
$\Ds\gamma$ has not yet been observed. Since the total width of the
$\Dsone(2536)$ is below the MeV scale, a suppressed radiative branching
fraction is experimentally challenging to measure
\cite{BaBar:2011zpy}. {Very recently, the LHCb Collaboration performed a dedicated study of
muon-tagged $\Dsone(2460)^+$ and $\Dsone(2536)^+$ decays and obtained the
upper limit
$\Gamma_{\rm tot}[\Dsone(2460)^+]<1.28~\MeV$
at $90\%$ confidence level, adding new experimental information on these
axial-vector charm-strange states
\cite{LHCb:2026muontagged}.}

In the bottom-strange sector, the $\Bsone(5830)$ state has been observed in
$B^{\ast}K$ spectroscopy \cite{LHCb:2012kxf,CMS:2018qdl}, whereas its
radiative transition to $\Bs\gamma$ remains unknown. A second, lower
$J^P=1^+$ bottom-strange state is also expected theoretically as the
bottom partner of the $\Dsone(2460)$. Lattice-QCD calculations
\cite{Lang:2015hza}, hadronic-molecule studies \cite{Faessler:2008vc}, and
coupled-channel analyses \cite{Yang:2022vdb} place this state below or near
the $B^{\ast}K$ threshold, with mass estimates in the range
$5.75$--$5.78~\mathrm{GeV}$. {This lower state has not yet been
observed experimentally. In the present work, we denote it by
$B_{s1}(5750)$ and adopt the lattice-QCD mass prediction of
Ref.~\cite{Lang:2015hza}.}

Radiative transitions of heavy-light mesons have been studied using
relativistic and relativized quark models
\cite{Goity:2000dk,Godfrey:2005ww,Godfrey:2016nwn,Chen:2020ejk},
potential and nonrelativistic quark models
\cite{Radford:2009bs,Lu:2016bbk,Green:2016occ,Li:2021qgz},
an unquenched coupled-channel model
\cite{Zhang:2024zll},
{vector-meson-dominance (VMD) approaches
\cite{ColangeloRef10,ColangeloRef16},}
effective-field-theory and hadronic-molecule approaches
\cite{Faessler:2008vc,Cleven:2014oka,Fu:2021wde},
and light-cone QCD sum rules
\cite{Colangelo:2005hv,Pullin:2021ebn}.

More recently, phenomenological and potential-model analyses
\cite{Bondar:2023mua,Bondar:2025smw} have focused on the role of mixing in
the radiative decays of the axial-vector states. {Within
light-cone QCD sum rules, the $\Dsone(2460)\to\Ds\gamma$ transition is
predicted to be one of the dominant radiative decay modes of the
$\Dsone(2460)$~\cite{Colangelo:2005hv}.} Bondar and Milstein
\cite{Bondar:2023mua,Bondar:2025smw} argued that the large difference between
the $\Dsone(2460)\to\Ds\gamma$ and $\Dsone(2536)\to\Ds\gamma$ decay widths
originates from interference between the $^{1}P_{1}$ and $^{3}P_{1}$
components of the mixed physical states. The analysis was also extended to
the corresponding bottom-strange system, including both the observed
$\Bsone(5830)$ and the predicted lower $1^+$ state.

An important question is whether this interference mechanism also emerges
{within the light-cone QCD sum-rule framework.} In a quark-model description,
the cancellation can be traced directly to the components of the meson wave
function. In light-cone QCD sum rules, the corresponding effect must instead
arise from the correlation functions constructed with independent
interpolating currents for the $^{1}P_{1}$ and $^{3}P_{1}$ configurations.
It is therefore interesting to examine whether the sum-rule calculation
reproduces the same interference pattern.

In this work, we calculate the radiative decays
$\Dsone(2460)\to\Ds\gamma$,
$\Dsone(2536)\to\Ds\gamma$,
$B_{s1}(5750)\to\Bs\gamma$, and
$\Bsone(5830)\to\Bs\gamma$
{within the light-cone QCD sum-rule framework.} We retain the strange-quark-mass
contributions and employ photon distribution amplitudes up to twist four. The
physical axial-vector states are constructed from two independent
interpolating currents through the $^{1}P_{1}$--$^{3}P_{1}$ mixing scheme.
This allows us to study directly how the two basis-current contributions interfere in the physical radiative amplitudes. In the
bottom-strange sector, the lower $B_{s1}(5750)$ state is included as a
theoretical benchmark, while the result for the experimentally established
$\Bsone(5830)$ constitutes a prediction for its radiative decay.

The paper is organized as follows. In Sec.~\ref{sec:framework}, we introduce
the interpolating currents and derive the light-cone QCD sum rules for the
transition form factors. In Sec.~\ref{sec:numerics}, we present the numerical
analysis, compare our results with previous theoretical studies, and discuss
the current experimental status. Our conclusions are summarized in
Sec.~\ref{sec:concl}.

\section{Theoretical framework}
\label{sec:framework}

We consider the radiative transitions
\begin{equation}
  A_i(p',\eta)\to P(p)\gamma(q,\varepsilon),
  \qquad p'=p+q,
  \qquad i=1,2,
\end{equation}
where $A_i$ is a physical heavy axial-vector meson and $P$ is the
corresponding pseudoscalar meson. Here, $\eta$ and $\varepsilon$ denote the
polarization vectors of the axial-vector meson and the photon, respectively.
The heavy quark is denoted by $Q=c,b$.
{Since all channels considered here are strange, the light-quark
field is denoted by $s$ throughout, while $q$ is reserved exclusively for the
photon momentum. Axial-vector and pseudoscalar mesons masses are denoted by $m_{A_i}$ and $m_P$, respectively.}

The transition is described by the following vacuum correlation function:
\begin{equation}
\Pi_{\mu\nu}^{(i)}(p',q)
=
i^2
\int d^4x
\int d^4y\,
e^{ip\cdot x+iq\cdot y}
\langle 0|
T
\left\{
J_\mu^{(i)}(x)
J_\nu^{\rm em}(y)
J_5^\dagger(0)
\right\}
|0\rangle ,
\label{corr}
\end{equation}
where
\begin{equation}
J_\nu^{\rm em}
=
{e_s\,\bar s\gamma_\nu s}
+
e_Q\,\bar Q\gamma_\nu Q
\end{equation}
is the electromagnetic current, with {$e_s$} and $e_Q$ denoting the
electric charges of the strange and heavy quarks, respectively. The
pseudoscalar meson is interpolated by the current
\begin{equation}
{J_5=\bar s\,i\gamma_5Q.}
\end{equation}

For the axial-vector channel, we employ two independent interpolating
currents:
\begin{equation}
\begin{aligned}
J_\mu^A
&=
\bar s\gamma_\mu\gamma_5Q,
\\
J_\mu^B
&=
\frac{i}{m_Q+m_s}
\bar s
\sigma_{\mu\nu}\gamma_5Q
\,p'^{\,\nu}.
\end{aligned}
\label{basiscurrents}
\end{equation}
{We identify $J_\mu^A$ with the $^{3}P_1$ basis state and
$J_\mu^B$ with the $^{1}P_1$ basis state, adopting the same relative phase
convention as in Ref.~\cite{Aliev:2025mixing}. With this convention, the
physical states are defined as}
\begin{equation}
\begin{aligned}
|A_1\rangle
&=
\sin\theta\,|{}^3P_1\rangle
+
\cos\theta\,|{}^1P_1\rangle,
\\
|A_2\rangle
&=
\cos\theta\,|{}^3P_1\rangle
-
\sin\theta\,|{}^1P_1\rangle.
\end{aligned}
\label{eq:state-mixing-convention}
\end{equation}
{Consistent with the state-mixing convention in
Eq.~\eqref{eq:state-mixing-convention}, we construct the interpolating
currents for the two physical states using the same rotation:}
\begin{equation}
\begin{pmatrix}
J_\mu^{(1)}\\[2pt]
J_\mu^{(2)}
\end{pmatrix}
=
\begin{pmatrix}
\sin\theta & \cos\theta\\
\cos\theta & -\sin\theta
\end{pmatrix}
\begin{pmatrix}
J_\mu^A\\[2pt]
J_\mu^B
\end{pmatrix}.
\label{mixingcurrents}
\end{equation}
Following the standard QCD sum-rule procedure, the correlation function is
calculated in two different representations. On the hadronic side, it is
expressed in terms of physical hadronic states, whereas on the QCD side it is
evaluated in terms of quark and gluon degrees of freedom.

The hadronic representation is obtained by inserting complete sets of states
with the same quantum numbers as the interpolating currents and isolating the
ground-state contributions. This gives
\begin{equation}
\Pi_{\mu\nu}^{(i)}
=
\frac{
\langle 0|J_\mu^{(i)}|A_i(p',\eta)\rangle
\langle A_i(p',\eta)|J_\nu^{\rm em}|P(p)\rangle
\langle P(p)|J_5^\dagger|0\rangle
}
{
{(m_{A_i}^2-p'^2)}
(m_P^2-p^2)
}
+\cdots ,
\label{eq:hadronic-pole}
\end{equation}
where the ellipsis denotes contributions from higher resonances and continuum
states.

The relevant matrix elements are defined as
\begin{equation}
\langle 0|J_\mu^{(i)}|A_i(p',\eta)\rangle
=
{m_{A_i}f_i\eta_\mu} ,
\end{equation}
and
\begin{equation}
\langle P(p)|J_5^\dagger|0\rangle
=
{\frac{m_P^2f_P}{m_Q+m_s}.}
\end{equation}
Here, $f_i$ and $f_P$ denote the decay constants of the axial-vector meson
$A_i$ and the pseudoscalar meson $P$, respectively.

Lorentz covariance and electromagnetic gauge invariance allow the matrix element to be parametrized as
\begin{equation}
\langle A_i(p',\eta)
|
J_\nu^{\rm em}
|
P(p)
\rangle
=
 g_i
\left[
(p\!\cdot\!q)\eta_\nu
-
(\eta\!\cdot\!q)p_\nu
\right],
\label{matrixelement}
\end{equation}
where
\begin{equation*}
q=p'-p,
\end{equation*}
and $g_i(q^2)$ denotes the electric dipole transition form factor. Our aim is
to determine this form factor for a real photon, $q^2=0$.

{After summing over the axial-vector polarization states,
Eq.~\eqref{eq:hadronic-pole} takes the form}
\begin{equation}
\Pi_{\mu\nu}^{(i)}
=
\frac{
{m_{A_i} f_i}
g_i
m_P^2 f_P
}
{
{(m_Q+m_s)}
{(m_{A_i}^2-p'^2)}
(m_P^2-p^2)
}
\left[
(p\!\cdot\!q)
\left(
-g_{\mu\nu}
+
{\frac{p'_\mu p'_\nu}{m_{A_i}^2}}
\right)
-
p_\nu
\left(
-q_\mu
+
\frac{p'_\mu(p'\!\cdot\!q)}
{m_{A_i}^2}
\right)
\right].
\label{hadronic}
\end{equation}

After contracting both sides with the photon polarization vector
$\varepsilon^\nu$, the correlation function can be written as
\begin{equation}
\Pi_{\mu\nu}^{(i)}
\varepsilon^\nu
=
\frac{
{m_{A_i} f_i}
g_i
m_P^2 f_P
}
{
{(m_Q+m_s)}
{(m_{A_i}^2-p'^2)}
(m_P^2-p^2)
}
\left\{
-\varepsilon_\mu(p\!\cdot\!q)
+
(p\!\cdot\!\varepsilon)q_\mu
+\cdots
\right\},
\end{equation}
where the omitted terms correspond to other independent Lorentz structures.
We derive the light-cone QCD sum rules for the transition form factors
$g_i(q^2=0)$ from the coefficient of the Lorentz structure
$(p\!\cdot\!\varepsilon)q_\mu-\varepsilon_\mu(p\!\cdot\!q)$.


We now turn to the calculation of the correlation function on the QCD side.
The QCD representation is evaluated in the deep Euclidean region,
\begin{equation*}
p^2\ll {(m_Q+m_s)^2},
\qquad
p'^2\ll {(m_Q+m_s)^2},
\end{equation*}
where the operator product expansion (OPE) is applicable. To describe the
interaction with the external photon, we employ the background-field method,
in which the electromagnetic field is treated as a classical external field.

For a real photon with momentum $q$ and polarization vector
$\varepsilon_\mu$, the electromagnetic field-strength tensor is taken as
\begin{equation}
F_{\mu\nu}(x)
=
i
\left(
\varepsilon_\nu q_\mu
-
\varepsilon_\mu q_\nu
\right)
e^{iq\cdot x}.
\end{equation}

Within this formalism, the correlation function in Eq.~\eqref{corr} can be
rewritten as
\begin{equation}
\varepsilon^\nu \Pi_{\mu\nu}^{(i)}(p,q)
=
i
\int d^4x\,
e^{ip\cdot x}
\left\langle
0
\left|
T
\left\{
J_\mu^{(i)}(x)
J_5^\dagger(0)
\right\}
\right|
0
\right\rangle_F,
\label{backgroundcorr}
\end{equation}
where the subscript $F$ indicates that the vacuum expectation value is
evaluated in the presence of the external electromagnetic field.

The correlation function receives two types of contributions. The first
corresponds to the perturbative emission of the photon from one of the quark
propagators. The second describes the long-distance interaction of the photon
with the quark fields and is parameterized in terms of photon light-cone
distribution amplitudes (DAs).

Applying Wick's theorem, the correlation function can be expressed in terms of
the heavy- and light-quark propagators as
\begin{equation}
\varepsilon^\nu \Pi_{\mu\nu}^{(i)} =
i \int d^4x\,
e^{ip\cdot x}
\mathrm{Tr}
\left[
\Gamma_\mu^{(i)}
S_Q^{ab}(x)
\gamma_5
{S_s^{ba}(-x)}
\right]_F,
\label{wick}
\end{equation}
where $\Gamma_\mu^{(i)}$ denotes the Dirac structure associated with the
chosen interpolating current, while $a$ and $b$ are color indices.


It follows from Eq.~\eqref{wick} that the calculation of the correlation
function requires the light- and heavy-quark propagators in the presence of
external gluonic and electromagnetic background fields. {For
$j=s,Q$, the corresponding propagator can be written as}
\begin{align}
S_j(x)
=&
\int
\frac{d^4k}{(2\pi)^4}
e^{-ik\cdot x}
\frac{i(\slashed{k}+m_j)}
{k^2-m_j^2}
\nonumber\\
&
-i g_s
\int
\frac{d^4k}{(2\pi)^4}
e^{-ik\cdot x}
\int_0^1du
\left[
\frac{\slashed{k}+m_j}
{2(m_j^2-k^2)^2}
G^{\mu\nu}(ux)
\sigma_{\mu\nu}
+
\frac{u x_\mu
G^{\mu\nu}(ux)
\gamma_\nu}
{m_j^2-k^2}
\right]
\nonumber\\
&
-i e_j
\int
\frac{d^4k}{(2\pi)^4}
e^{-ik\cdot x}
\int_0^1du
\left[
\frac{\slashed{k}+m_j}
{2(m_j^2-k^2)^2}
F^{\mu\nu}(ux)
\sigma_{\mu\nu}
+
\frac{u x_\mu
F^{\mu\nu}(ux)
\gamma_\nu}
{m_j^2-k^2}
\right]
+\cdots ,
\label{fullprop}
\end{align}
{Here the subscript $j$ denotes either the strange quark ($s$) or
the heavy quark ($Q$).}
Furthermore, $G_{\mu\nu}$ and $F_{\mu\nu}$ are the external gluonic and
electromagnetic field-strength tensors, respectively, and $e_j$ is the
electric charge of the corresponding quark.

The nonperturbative contribution is obtained by replacing the light-quark
propagator in Eq.~\eqref{wick} according to
\begin{equation}
S_{\alpha\beta}(x)
=
-\frac{1}{4}
\left(\Gamma_k\right)_{\alpha\beta}
\,
\bar s(x)\Gamma_k s(0),
\label{nonpertprop}
\end{equation}
where
\begin{equation}
\Gamma_k
=
\left\{
I,\,
\gamma_5,\,
\gamma_\mu,\,
i\gamma_\mu\gamma_5,\,
\frac{\sigma_{\mu\nu}}{\sqrt{2}}
\right\}
\end{equation}
forms a complete basis of Dirac matrices.


This replacement gives rise to the following nonlocal matrix elements:
\begin{align}
\langle
0
|
\bar s(x)
\Gamma_i
s(0)
|
0
\rangle_F,
\nonumber\\
\langle
0
|
\bar s(x)
\Gamma_i
G_{\mu\nu}
s(0)
|
0
\rangle_F,
\nonumber\\
\langle
0
|
\bar s(x)
\Gamma_i
F_{\mu\nu}
s(0)
|
0
\rangle_F,
\end{align}
which are parameterized in terms of photon light-cone distribution amplitudes
of definite twist. In the present analysis, we employ the complete set of
two-particle and three-particle photon distribution amplitudes up to twist
four. Their explicit expressions are collected
in Appendix~\ref{app:sumrule-expressions}.

The OPE representation of the correlation function can therefore be written
as the sum of perturbative and nonperturbative contributions:
\begin{equation}
\Pi_K^{\mathrm{OPE}}
=
\Pi_K^{\mathrm{pert}}
+
\Pi_K^{\mathrm{2p}}
+
\Pi_K^{\mathrm{3p},g}
+
\Pi_K^{\mathrm{3p},\gamma},
\qquad K=A,B,
\label{decomp}
\end{equation}
where $\Pi_K^{\mathrm{2p}}$ denotes the contribution of two-particle
quark--antiquark photon DAs, while $\Pi_K^{\mathrm{3p},g}$ and
$\Pi_K^{\mathrm{3p},\gamma}$ denote the three-particle contributions
involving quark--antiquark--gluon and quark--antiquark--photon operators,
respectively. The two-particle contribution contains terms of twist two,
three, and four. Each term in Eq.~\eqref{decomp} refers to the invariant
amplitude multiplying the gauge-invariant Lorentz structure
\begin{equation*}
(p\!\cdot\!\varepsilon)q_\mu
-
\varepsilon_\mu(p\!\cdot\!q).
\end{equation*}

The perturbative contribution is written in terms of spectral densities, and
the continuum subtraction is implemented using quark--hadron duality. Its
double-Borel-transformed expression is given below in terms of
$\rho_A^P(s)$ and $\rho_B^P(s)$. The two- and three-particle photon-DA
contributions are transformed separately using the continuum-subtraction
prescription specified below.

Since the physical interpolating currents are linear combinations of the two
basis currents, the corresponding invariant amplitudes are given by
\begin{equation}
\begin{aligned}
\widehat{\Pi}_1
&=
\sin\theta\,
\widehat{\Pi}_A
+
\cos\theta\,
\widehat{\Pi}_B,
\\
\widehat{\Pi}_2
&=
\cos\theta\,
\widehat{\Pi}_A
-
\sin\theta\,
\widehat{\Pi}_B.
\end{aligned}
\label{mixingPi}
\end{equation}

{
Having determined the coefficient of the selected gauge-invariant Lorentz
structure on both the hadronic and OPE sides, we equate the two
representations. We then perform double Borel transformations with respect to
$p'^2$ and $p^2$, which suppress contributions from higher states, and
implement the continuum subtraction using quark--hadron duality. Combining
the Borel-transformed basis amplitudes with the physical-current rotation
yields the sum rules for the transition form factors $g_i(0)$.
}
{
Before imposing a relation between the two Borel parameters, the matching
condition takes the form
\begin{equation}
g_i(0)\,m_{A_i} f_i
\frac{m_P^2f_P}{m_Q+m_s}
\exp\!\left(
-\frac{m_{A_i}^2}{M_1^2}
-\frac{m_P^2}{M_2^2}
\right)
=
\widehat{\Pi}_i(M_1^2,M_2^2,s_0),
\label{eq:general-sum-rule}
\end{equation}
It is convenient to introduce the effective Borel parameter and momentum
fractions
\begin{equation}
M^2
=
\frac{M_1^2M_2^2}{M_1^2+M_2^2},
\qquad
u_0
=
\frac{M_2^2}{M_1^2+M_2^2},
\qquad
a_0
=
1-u_0.
\label{eq:borel-definitions}
\end{equation}
In the numerical analysis, we employ the symmetric double-Borel prescription
\begin{equation}
M_1^2=M_2^2=2M^2,
\qquad
u_0=a_0=\frac{1}{2}.
\label{eq:symmetric-borel}
\end{equation}
With this choice, Eq.~\eqref{eq:general-sum-rule} becomes
\begin{equation}
g_i(0)\,m_{A_i} f_i
\frac{m_P^2f_P}{m_Q+m_s}
\exp\!\left[
-\frac{m_{A_i}^2+m_P^2}{2M^2}
\right]
=
\widehat{\Pi}_i(M^2,s_0).
\label{eq:symmetric-sum-rule}
\end{equation}
}


After the double Borel transformation, each invariant amplitude is decomposed
into perturbative, twist-2, twist-3, twist-4, and three-particle gluonic and
electromagnetic contributions:
\begin{equation}
\begin{aligned}
\widehat{\Pi}_{A(B)}
={}&
\widehat{\Pi}_{A(B)}^{\rm pert}
+
\widehat{\Pi}_{A(B)}^{\rm twist2}
+
\widehat{\Pi}_{A(B)}^{\rm twist3}
+
\widehat{\Pi}_{A(B)}^{\rm twist4}
+
\widehat{\Pi}_{A(B)}^{\rm 3p,g}
+
\widehat{\Pi}_{A(B)}^{\rm 3p,\gamma}.
\end{aligned}
\label{decomposition}
\end{equation}
{The perturbative contributions are expressed in terms of the
spectral densities $\rho_A(s)$ and $\rho_B(s)$ associated with the
axial-vector and tensor basis currents, respectively. }
\begingroup

For photon emission from a quark of flavor $a$, with the other quark denoted
by $b$, we introduce the auxiliary spectral density
\begin{align}
\rho_a(s;m_a,m_b,e_a)
={}&
-\frac{3e_a}{8\pi^2}
\Bigg[
2m_a
\ln\!\left(
\frac{
s-m_b^2+m_a^2-\sqrt{\lambda(s,m_b^2,m_a^2)}
}{
s-m_b^2+m_a^2+\sqrt{\lambda(s,m_b^2,m_a^2)}
}
\right)
\notag\\
&\qquad+
(m_b-m_a)
\frac{m_b^2-m_a^2-s}{s^2}
\sqrt{\lambda(s,m_b^2,m_a^2)}
\Bigg].
\label{eq:rho-emission}
\end{align}
The perturbative spectral density associated with the axial-vector basis
current is then
\begin{equation}
\rho_A(s)
=
\rho_s(s;m_s,m_Q,e_s)
-
\rho_Q(s;m_Q,m_s,e_Q).
\label{eq:rhoA-pert}
\end{equation}
\endgroup

{For the tensor basis current, the perturbative spectral density is}
\begin{align}
\rho_B(s)
&=
-\frac{3}{8\pi^2}
\Bigg\{
e_s
\Bigg[
\frac{2m_Qm_s}{m_Q+m_s}
\ln\!\left(
\frac{
s-m_Q^2+m_s^2-\sqrt{\lambda(s,m_Q^2,m_s^2)}
}{
s-m_Q^2+m_s^2+\sqrt{\lambda(s,m_Q^2,m_s^2)}
}
\right)
\notag\\
&\qquad+
\frac{s}{m_Q+m_s}
\frac{m_Q^2-m_s^2-s}{s^2}
\sqrt{\lambda(s,m_Q^2,m_s^2)}
\Bigg]
\notag\\
&\qquad-
e_Q
\Bigg[
\frac{2\left(2m_Q^2-m_Qm_s\right)}{m_Q+m_s}
\ln\!\left(
\frac{
s-m_s^2+m_Q^2-\sqrt{\lambda(s,m_s^2,m_Q^2)}
}{
s-m_s^2+m_Q^2+\sqrt{\lambda(s,m_s^2,m_Q^2)}
}
\right)
\notag\\
&\qquad\quad-
\frac{s}{m_Q+m_s}
\frac{m_s^2-m_Q^2-s}{s^2}
\sqrt{\lambda(s,m_s^2,m_Q^2)}
\Bigg]
\Bigg\},
\label{eq:rhoB-pert}
\end{align}

\begingroup

where 
\begin{equation}
\lambda(a,b,c)
=
(a-b-c)^2-4bc,
\label{eq:kallen}
\end{equation}
\endgroup
is the K\"all\'en function.

{The corresponding continuum-subtracted and Borel-transformed
perturbative amplitudes are}
\begingroup

\begin{equation}
\widehat{\Pi}_K^{\rm pert}(M^2,s_0)
=
\int_{(m_Q+m_s)^2}^{s_0}
ds\,
e^{-s/M^2}
\rho_K(s),
\qquad
K=A,B.
\label{eq:perturbative-amplitudes}
\end{equation}
\endgroup
{ 
For the photon-DA contributions, the continuum subtraction is implemented
term by term. The twist-2, twist-3, two-particle twist-4, gluonic
three-particle, and electromagnetic three-particle contributions are all
multiplied by
\begin{equation*}
e^{-m_Q^2/M^2}-e^{-s_0/M^2}.
\end{equation*}
The finite strange-quark mass does not modify this soft-sector continuum
factor. It is retained in the perturbative threshold $(m_Q+m_s)^2$, the
spectral densities, the current normalization, and the mass-dependent
coefficient functions.
}

{With this prescription, the two-particle contributions associated with the
axial-vector basis current are}
\begin{align}
\widehat{\Pi}_A^{\rm twist2}
&=
{e_s f_{\gamma,s}^{\perp}}
\left(
e^{-m_Q^2/M^2}
-
e^{-s_0/M^2}
\right)
M^2\phi_\gamma(u_0),
\notag\\
\widehat{\Pi}_A^{\rm twist3}
&=
e_s f_{3\gamma}m_Q
\left(
e^{-m_Q^2/M^2}
-
e^{-s_0/M^2}
\right)
{\psi^{(V)}(u_0)},
\notag\\
\widehat{\Pi}_A^{\rm twist4}
&=
\frac{e_s\ssbar}{4} \left(
e^{-m_Q^2/M^2}
-
e^{-s_0/M^2}
\right)
\left[
\mathcal A(u_0)
+
2\mathcal B(u_0)
\right]
\left(
1+\frac{m_Q^2}{M^2}
\right).
\label{eq:piA-components}
\end{align}

{Here, $\phi_\gamma(u)$, $\psi^{(V)}(u)$, $\mathcal A(u)$, and
$\mathcal B(u)$ denote the two-particle photon DAs of twist two, three, and
four, respectively. Their explicit expressions and normalization conventions
are collected in Appendix~\ref{app:sumrule-expressions}.}

{For the tensor basis current, the twist-3 contribution is}
\begingroup

\begin{equation}
\widehat{\Pi}_B^{\rm twist3}
=
\frac{
e_s f_{3\gamma}
}{
m_Q+m_s
}
\left(
e^{-m_Q^2/M^2}
-
e^{-s_0/M^2}
\right)
\left[
m_Q^2{\psi^{(V)}(u_0)}
-
\left.
\frac{M^2}{2}
\frac{d}{da}
\left\{
(1-a){\psi^{(V)}(a)}
\right\}
\right|_{a=u_0}
\right].
\label{eq:piB-twist3}
\end{equation}
\endgroup

{The twist-2 and twist-4 contributions associated with the tensor
basis current} are related to the corresponding axial-vector-current
contributions through
\begin{equation}
\widehat{\Pi}_B^{\rm twist2}
=
\frac{m_Q}{m_Q+m_s}
\widehat{\Pi}_A^{\rm twist2},
\qquad
\widehat{\Pi}_B^{\rm twist4}
=
\frac{m_Q}{m_Q+m_s}
\widehat{\Pi}_A^{\rm twist4}.
\label{eq:piB-twist-relations}
\end{equation}

{Explicit expressions for $\widehat{\Pi}_K^{\rm 3p,g}$ and
$\widehat{\Pi}_K^{\rm 3p,\gamma}$, with $K=A,B$, are collected in
Appendix~\ref{app:basis-amplitudes}.}
%

The sum rule in Eq.~\eqref{eq:general-sum-rule} depends on the decay constants
of the physical axial-vector mesons. These quantities are determined from the
two-point correlation function
\begin{equation}
\Pi_{\mu\nu}^{(i)}(p)
=
i
\int d^4x\,
e^{ip\cdot x}
\langle 0|
T
\left\{
J_\mu^{(i)}(x)
J_\nu^{(i)\dagger}(0)
\right\}
|0\rangle,
\label{twopoint}
\end{equation}
where the interpolating currents $J_\mu^{(i)}$ are defined in
Eq.~\eqref{mixingcurrents}.

The resulting sum rules for the decay constants are
\begin{align}
{f_1^2m_{A_1}^2e^{-m_{A_1}^2/M^2}}
&=
\sin^2\theta\,\widehat{\Pi}_{AA}
+
2\sin\theta\cos\theta\,\widehat{\Pi}_{AB}
+
\cos^2\theta\,\widehat{\Pi}_{BB},
\label{f1}\\
{f_2^2m_{A_2}^2e^{-m_{A_2}^2/M^2}}
&=
\cos^2\theta\,\widehat{\Pi}_{AA}
-
2\sin\theta\cos\theta\,\widehat{\Pi}_{AB}
+
\sin^2\theta\,\widehat{\Pi}_{BB}.
\label{f2}
\end{align}
Here,
$\widehat{\Pi}_{AA}$,
$\widehat{\Pi}_{AB}$,
and
$\widehat{\Pi}_{BB}$
denote the Borel-transformed invariant amplitudes obtained from the
corresponding two-point correlation functions, and their expressions are
presented in Appendix C.

{ 
Substituting the residues obtained from the two-point sum rules into the
symmetric transition sum rule, and using the physical-current rotation, we
obtain the form factors
\begin{align}
g_1(0)
={}&
\frac{m_Q+m_s}
{m_{A_1}f_1m_P^2f_P}
\exp\!\left(
\frac{m_{A_1}^2+m_P^2}{2M^2}
\right)
\left[
\sin\theta\,\widehat{\Pi}_A(M^2,s_0)
+
\cos\theta\,\widehat{\Pi}_B(M^2,s_0)
\right],
\notag\\
g_2(0)
={}&
\frac{m_Q+m_s}
{m_{A_2}f_2m_P^2f_P}
\exp\!\left(
\frac{m_{A_2}^2+m_P^2}{2M^2}
\right)
\left[
\cos\theta\,\widehat{\Pi}_A(M^2,s_0)
-
\sin\theta\,\widehat{\Pi}_B(M^2,s_0)
\right].
\label{eq:transition-form-factors}
\end{align}
For each transition, the invariant amplitudes in
Eq.~\eqref{eq:transition-form-factors} are evaluated using the corresponding
heavy-quark flavor, continuum threshold, and Borel window.
}
The transition form factors obtained from the light-cone QCD sum rules can be
used to determine the radiative decay widths of the physical axial-vector
mesons.
The partial decay width of the $A_i\to P\gamma$ process is
\begin{equation}
\Gamma(A_i\to P\gamma)
=
\frac{\alpha_{\mathrm{em}}}{3}
g_i^2(0)
\left(
{\frac{m_{A_i}^2-m_P^2}
{2m_{A_i}}}
\right)^3,
\label{eq:width}
\end{equation}
where $\alpha_{\mathrm{em}}$ is the electromagnetic fine-structure constant.

Once the transition form factors are determined, Eq.~\eqref{eq:width} gives
the corresponding radiative decay widths. The numerical analysis is presented
in the following section.

\section{Numerical analysis}
\label{sec:numerics}

The external numerical inputs required for the analysis, including the quark
and meson masses, pseudoscalar decay constants, condensates, and the
leading-twist photon-DA normalization, are collected in
Table~\ref{tab:inputs}. The remaining photon-DA parameters are listed in
Table~\ref{tab:app-photon-params}. For the leading-twist photon DA, we use the recent lattice determination
$f_{\gamma,s}^{\perp}$~\cite{Bacchio:2024pij}, which provides a more precise
input than the conventional susceptibility parameter
$\chi_s$~\cite{Rohrwild:2007yt}.

\begin{table}[t]
\centering
\caption{{The values of the input parameters used in the numerical
analysis. The condensate inputs and photon-DA normalization are quoted at
$\mu=1~\GeV$.}}
\label{tab:inputs}
\setlength{\tabcolsep}{6pt}
\renewcommand{\arraystretch}{1.15}
\begin{ruledtabular}
\begin{tabular}{ll}
Parameter & Value \\
\colrule
$\overline m_c(\overline m_c),\ \overline m_b(\overline m_b),\
 \overline m_s(2~\GeV)$
& $1.27(2),\ 4.18(3),\ 0.093(11)~\GeV$
\cite{ParticleDataGroup:2026} \\
{$m_{D_{s1}(2460)},\ m_{D_{s1}(2536)},\ m_{D_s}$}
& {$2.4595(6),\ 2.53512(6),\ 1.96835(7)~\GeV$}
\cite{ParticleDataGroup:2026} \\
{$m_{B_{s1}(5750)},\ m_{B_{s1}(5830)},\ m_{B_s}$}
& {$5.750(26)~\GeV$~\cite{Lang:2015hza},\
$5.82873(20),\ 5.36692(10)~\GeV$
\cite{ParticleDataGroup:2026}} \\
$f_{D_s},\ f_{B_s}$
& {$0.2499(5),\ 0.2303(13)~\GeV$}
\cite{FlavourLatticeAveragingGroupFLAG:2024oxs} \\
{$\langle\bar q q\rangle$}
& {$-(0.240\pm0.010)^3~\GeV^3$
\cite{Rohrwild:2007yt}} \\
{$
\langle\bar s s\rangle/\langle\bar q q\rangle$}
& {$0.80\pm0.10$
\cite{Colangelo:2005hv}} \\
{$m_0^2\equiv
\langle\bar s g_s\sigma\!\cdot\!G s\rangle/
\langle\bar s s\rangle$}
& {$0.80\pm0.20~\GeV^2$
\cite{Colangelo:2005hv}} \\
$f_{\gamma,s}^{\perp} \equiv \chi_s \langle \bar s s \rangle$
& $-55.1(1.9)~\MeV$~\cite{Bacchio:2024pij} \\
\end{tabular}
\end{ruledtabular}
\end{table}

{For the numerical analysis, we use the mixing angles obtained in
Ref.~\cite{Aliev:2025mixing},
\begin{equation}
\theta_{D_s}=26.6^\circ\pm0.6^\circ,
\qquad
\theta_{B_s}=38.5^\circ\pm0.1^\circ.
\label{eq:mixing-angles}
\end{equation}
Using these values in the two-point sum rules in Eqs.~\eqref{f1} and
\eqref{f2}, we obtain the physical axial-vector residues
$f_1^{(D_s)}=405^{+25}_{-26}~\MeV$ for $\Dsone(2460)$,
$f_2^{(D_s)}=168^{+7}_{-8}~\MeV$ for $\Dsone(2536)$,
$f_1^{(B_s)}=536^{+23}_{-26}~\MeV$ for $B_{s1}(5750)$, and
$f_2^{(B_s)}=89^{+4}_{-4}~\MeV$ for $\Bsone(5830)$. As a cross-check, at
$\theta=90^\circ$, the first physical current reduces to the unmixed
axial-vector current, and the corresponding results are compatible with
those reported in~\cite{Wang:2015mxa}.}

{Because the initial and final mesons in each channel are
relatively close in mass, we employ the symmetric Borel prescription
$M_1^2=M_2^2=2M^2$. The working regions are selected by requiring stability
of the transition form factors, convergence of the light-cone expansion, and
sufficient suppression of higher-state and continuum contributions. The
adopted intervals are listed in Table~\ref{tab:windows}.}

\begin{table}[t]
\centering
\caption{Borel windows and continuum-threshold intervals used in the
numerical analysis.}
\label{tab:windows}
\setlength{\tabcolsep}{8pt}
\renewcommand{\arraystretch}{1.15}
\begin{ruledtabular}
\begin{tabular}{lcc}
Channel & $M^2~(\GeV^2)$ & $s_0~(\GeV^2)$ \\
\colrule
$D_{s1}(2460)\to D_s\gamma$
& $3.0$--$4.5$ & $8.5$--$9.5$ \\
$D_{s1}(2536)\to D_s\gamma$
& $3.0$--$4.5$ & $9.0$--$10.0$ \\
$B_{s1}(5750)\to B_s\gamma$
& $10.0$--$14.0$ & $39.0$--$41.0$ \\
$B_{s1}(5830)\to B_s\gamma$
& $10.0$--$14.0$ & $40.0$--$42.0$ \\
\end{tabular}
\end{ruledtabular}
\end{table}

The heavy-quark-limit
value $\theta=35.3^\circ$ is retained as a reference.

\begin{figure}[t]
\centering
\includegraphics[width=\linewidth]{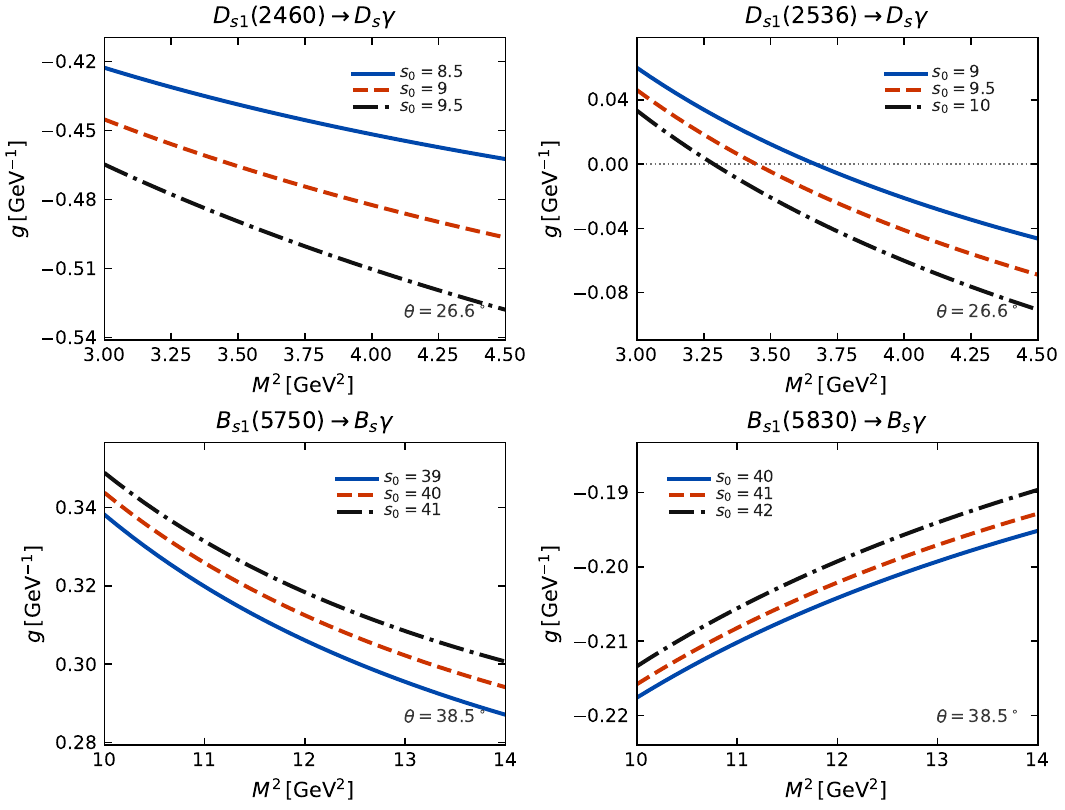}
\caption{Dependence of the transition form factors $g(0)$ on the Borel parameter
$M^2$ at fixed values of the continuum threshold $s_0$. The panels
show $D_{s1}(2460)\to D_s\gamma$, $D_{s1}(2536)\to D_s\gamma$,
$B_{s1}(5750)\to B_s\gamma$, and $B_{s1}(5830)\to B_s\gamma$.}
\label{fig:stability}
\end{figure}

{Figure~\ref{fig:stability} shows the dependence of the transition
form factors on $M^2$ at fixed values of $s_0$ and at the mixing angles given
in Eq.~\eqref{eq:mixing-angles}. The residual variation over the full $M^2$
and $s_0$ intervals is included in the uncertainty analysis.}

The zero crossing of $g_2(0)$ in the
$D_{s1}(2536)\to D_s\gamma$ panel of Fig.~\ref{fig:stability}
is a consequence of destructive interference between the two basis-current
contributions. The individual contributions vary smoothly with $M^2$ and
$s_0$, but enter the  amplitude with opposite signs and comparable
magnitudes. Their cancellation therefore leads to a zero whose position
changes with the variation of the input parameters. 

\begin{figure}[t]
\centering
\includegraphics[width=\linewidth]{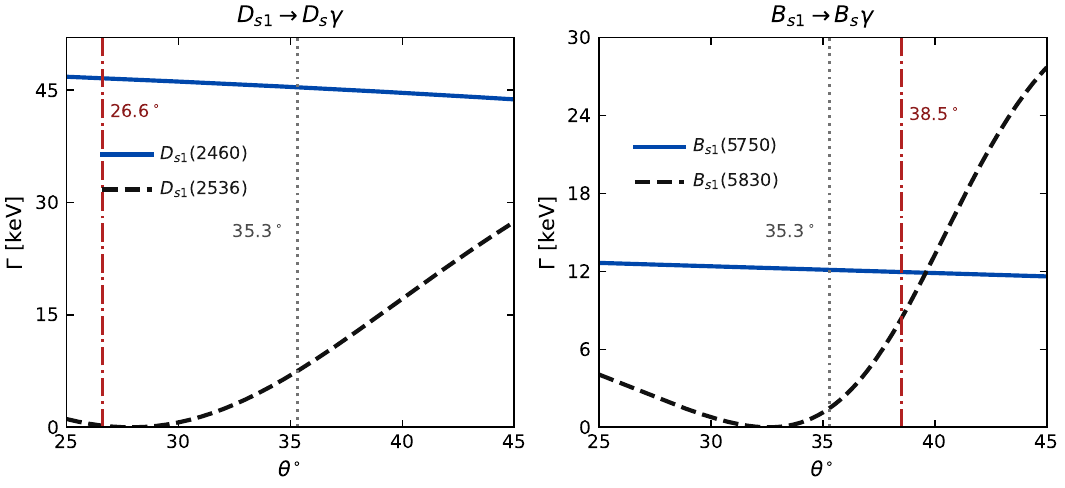}
\caption{Dependence of the radiative decay widths on the mixing angle. The
charm-strange and bottom-strange channels are shown in the left and right
panels, respectively. The mixing angles
$\theta_{D_s}=26.6^\circ$ and $\theta_{B_s}=38.5^\circ$ are indicated in the
corresponding panels, while $\theta=35.3^\circ$ denotes the heavy-quark-limit
reference value. The curves are evaluated with the remaining input parameters
fixed at their central values and are shown only to illustrate the
mixing-angle dependence; the final results are given in
Table~\ref{tab:results}.}
\label{fig:theta}
\end{figure}

{The  transition amplitudes are linear combinations of the
two basis-current contributions. When these contributions have opposite
signs, their destructive interference can produce a clear minimum in
the radiative width as the mixing angle is varied. This behavior is
illustrated in Fig.~\ref{fig:theta}. The remaining inputs are fixed at their
central values in this figure; the curves therefore illustrate only the
mixing-angle dependence and should not be identified with the final
results in Table~\ref{tab:results}.}

The uncertainties are obtained by varying the QCD and hadronic inputs within
their quoted ranges, together with the Borel parameter, continuum threshold,
and mixing angles. We quote the median values and the corresponding central
$68\%$ intervals. The resulting form factors and radiative widths are given
in Table~\ref{tab:results}.

The most notable feature is the strong suppression of
$D_{s1}(2536)\to D_s\gamma$ relative to
$D_{s1}(2460)\to D_s\gamma$. As seen from Figs.~\ref{fig:stability}
and~\ref{fig:theta}, this behavior originates from destructive interference
between the two basis-current contributions in the higher charm-strange
state, while the corresponding contributions add constructively for
$D_{s1}(2460)$. The calculation therefore reproduces within the light-cone
QCD sum-rule framework the cancellation mechanism discussed in
Refs.~\cite{Bondar:2023mua,Bondar:2025smw}. 

In the bottom-strange sector, the two radiative widths are comparable within
their uncertainties. The mixing-angle dependence of the higher state remains
significant, but the value of $\theta_{B_s}$ used in the numerical analysis
lies away from the strongest cancellation region. We emphasize that
$B_{s1}(5750)$ is not an experimentally established state and is included as
the theoretically expected lower $J^P=1^+$ bottom-strange configuration.

\begin{figure}[t]
\centering
\includegraphics[width=\linewidth]{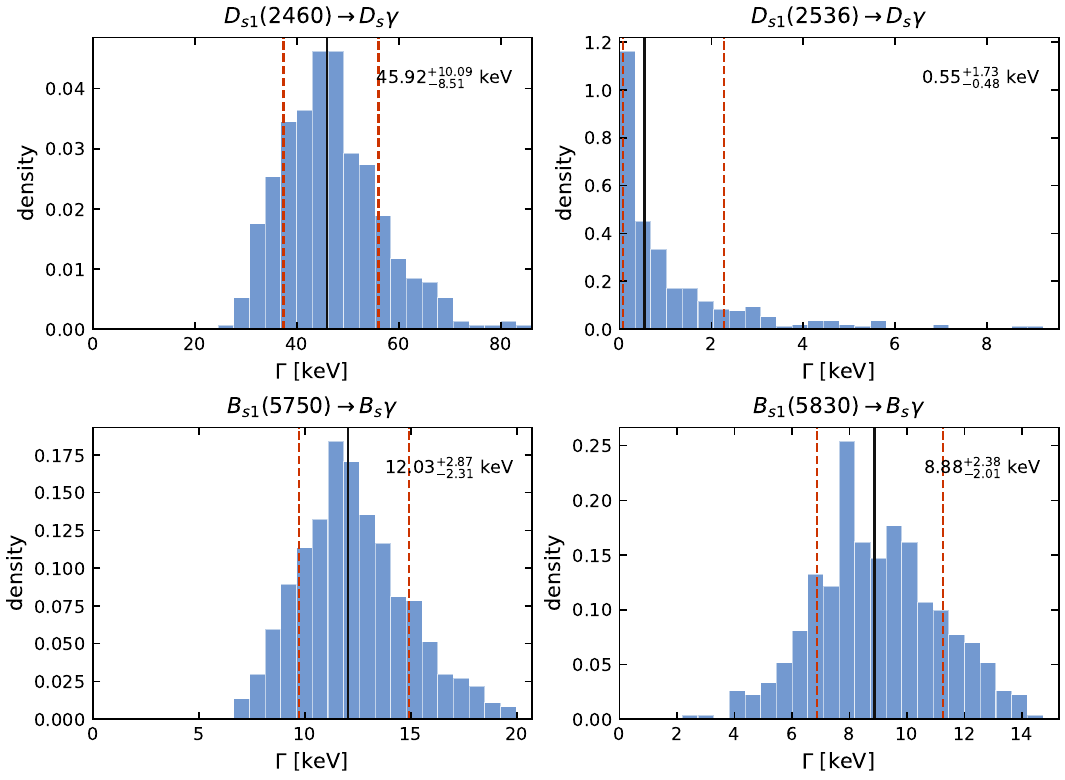}
\caption{Monte Carlo distributions of the radiative decay widths obtained
with the lattice-normalized photon input and the mixing angles
$\theta_{D_s}=26.6^\circ\pm0.6^\circ$ and
$\theta_{B_s}=38.5^\circ\pm0.1^\circ$. The remaining QCD and hadronic inputs,
together with the Borel parameter, continuum threshold, and two-point residue
samples, are varied as in the final uncertainty analysis. Solid vertical lines
indicate the medians, while dashed lines delimit the $16$th--$84$th percentile
intervals.}
\label{fig:mc}
\end{figure}

The width distributions shown in Fig.~\ref{fig:mc} reflect the different
sensitivities of the four channels to the input parameters. In particular,
the distribution for $D_{s1}(2536)\to D_s\gamma$ is strongly asymmetric,
consistent with its cancellation-sensitive amplitude.

Table~\ref{tab:comparison_literature} compares our results with theoretical predictions
available in the literature. The predictions differ considerably, especially
for channels in which the two mixed-state contributions partially cancel.
This sensitivity reflects the different treatments of the axial-vector states,
mixing, and transition amplitudes adopted in the various approaches. Despite
these differences in the absolute decay widths, the suppression of the higher
charm-strange state is also found in Ref.~\cite{Bondar:2025smw}.

Finally, combining our prediction for
$D_{s1}(2536)\to D_s\gamma$ with the measured total width gives
\begin{equation}
\mathcal B[D_{s1}(2536)\to D_s\gamma]
=
\left(0.60^{+1.86}_{-0.52}\right)\times10^{-3},
\label{eq:br-ds12536}
\end{equation}
where the uncertainty is dominated by that of the radiative amplitude. The small branching fraction makes the experimental observation of this decay mode challenging.

\begin{table}[t]
\centering
\caption{{The central values are medians, and the uncertainties
denote the central $68\%$ propagated uncertainty intervals.}}
\label{tab:results}
\setlength{\tabcolsep}{8pt}
\renewcommand{\arraystretch}{1.15}
\begin{ruledtabular}
\begin{tabular}{lcc}
Channel & $|g_i|~(\GeV^{-1})$ & $\Gamma~(\keV)$ \\
\colrule
$D_{s1}(2460)\to D_s\gamma$
& ${0.47_{-0.05}^{+0.05}}$
& ${45.9_{-8.5}^{+10.1}}$ \\
$D_{s1}(2536)\to D_s\gamma$
& ${0.04_{-0.03}^{+0.04}}$
& ${0.55_{-0.48}^{+1.73}}$ \\
$B_{s1}(5750)\to B_s\gamma$
& ${0.32_{-0.02}^{+0.02}}$
& ${12.0_{-2.3}^{+2.9}}$ \\
$B_{s1}(5830)\to B_s\gamma$
& ${0.20_{-0.02}^{+0.03}}$
& ${8.9_{-2.0}^{+2.4}}$ \\
\end{tabular}
\end{ruledtabular}
\end{table}

\begin{table*}[t]
\centering
\caption{Comparison with selected theoretical predictions. All widths are
given in keV.}
\label{tab:comparison_literature}
\renewcommand{\arraystretch}{1.20}
\begin{adjustbox}{max width=\textwidth}
\begin{tabular}{|l|c|c|c|c|}
\hline\hline
Approach / Ref.
& $D_{s1}(2460)\to D_s\gamma$
& $D_{s1}(2536)\to D_s\gamma$
& $B_{s1}(5750)\to B_s\gamma$
& $B_{s1}(5830)\to B_s\gamma$ \\
\hline
LCSR \cite{Colangelo:2005hv}
& $19$--$29$ & -- & -- & -- \\
VMD \cite{ColangeloRef10,ColangeloRef16}
& $3.3$ & -- & -- & -- \\
Relativized quark model \cite{Godfrey:2003kg,Godfrey:2005ww}
& $6.2$ & $15$ & -- & -- \\
Relativistic quark model \cite{Goity:2000dk}
& $10.3$--$17.2$ & $25.2$--$31.1$ & -- & -- \\
Potential model \cite{Green:2016occ}
& $13.2$ & $61.2$ & -- & -- \\
Potential model \cite{Radford:2009bs}
& $12.8$ & $54.5$ & -- & -- \\
Relativistic quark model \cite{Chen:2020ejk}
& $3.53$--$3.61$ & $18.18$--$18.85$ & -- & -- \\
Unquenched coupled-channel model \cite{Zhang:2024zll}
& $0.98$--$6.99$ & -- & $2.42$--$37.27$ & -- \\
Heavy-quark framework \cite{Korner:1992pz}
& -- & $1.6\pm2.3$ & -- & -- \\
Chiral quark model \cite{Li:2021qgz}
& -- & -- & $37$ & $27$ \\
Relativized quark model \cite{Godfrey:2016nwn}
& -- & -- & $70.6$ & $47.8$ \\
Constituent quark model \cite{Lu:2016bbk}
& -- & -- & $97.7$ & $56.6$ \\
Potential model \cite{Bondar:2025smw}
& $297$ & $12$ & $47$ & $28$ \\
This work (LCSR)
& ${45.9_{-8.5}^{+10.1}}$
& ${0.55_{-0.48}^{+1.73}}$
& ${12.0_{-2.3}^{+2.9}}$
& ${8.9_{-2.0}^{+2.4}}$ \\
\hline\hline
\end{tabular}
\end{adjustbox}

\end{table*}

\section{Conclusion}
\label{sec:concl}

We have studied the radiative decays
$\Dsone(2460)\to\Ds\gamma$,
$\Dsone(2536)\to\Ds\gamma$,
$B_{s1}(5750)\to\Bs\gamma$, and
$\Bsone(5830)\to\Bs\gamma$
within light-cone QCD sum rules, taking into account the
$^{1}P_{1}$--$^{3}P_{1}$ mixing of the physical axial-vector states.
The decay constants are determined consistently from two-point
sum rules and used in the transition analysis. To our knowledge, the
$\Dsone(2536)\to\Ds\gamma$ mode is studied here for the first time within
the light-cone QCD sum-rule framework.

The main result is the strong suppression of
$\Dsone(2536)\to\Ds\gamma$ relative to
$\Dsone(2460)\to\Ds\gamma$. The two basis-current contributions interfere
constructively for the lower state, whereas they nearly cancel for the
higher state. The light-cone QCD sum-rule calculation therefore provides an
independent realization of the interference mechanism discussed in
Refs.~\cite{Bondar:2023mua,Bondar:2025smw}. The resulting
$\Dsone(2536)\to\Ds\gamma$ width is at the sub-keV level, corresponding to a
branching fraction of order $10^{-3}$. This strong suppression is consistent
with the present difficulty of observing this radiative mode experimentally.

The bottom-strange system shows a different pattern. At the mixing angle used
in the numerical analysis, the predicted widths of
$B_{s1}(5750)\to\Bs\gamma$ and $\Bsone(5830)\to\Bs\gamma$ are comparable
within their uncertainties, although the higher-state amplitude remains
sensitive to the interference of the two basis components. The
$B_{s1}(5750)$ state is included as the theoretically expected lower
$J^P=1^+$ bottom-strange configuration and should not be interpreted as an
experimentally established resonance.

Our results show that the radiative decays of heavy-strange axial-vector
mesons are strongly influenced by the relative magnitude and sign of the
$^{1}P_{1}$ and $^{3}P_{1}$ contributions and, consequently, by the mixing
angle. This sensitivity is particularly strong for channels close to a
cancellation and also helps explain the wide spread among existing theoretical
predictions. Future measurements of
$\Dsone(2536)\to\Ds\gamma$ and $\Bsone(5830)\to\Bs\gamma$, together with
improved information on other radiative modes, could therefore provide
constraints on the $^{1}P_{1}$--$^{3}P_{1}$ mixing angles and discriminate
among different descriptions of the physical axial-vector states. Searches
for the lower bottom-strange $1^+$ state would provide an additional test of
this picture.


\appendix

\section{Photon distribution amplitudes }
\label{app:sumrule-expressions}

{Photon DAs provide the principal nonperturbative input to the
light-cone QCD sum rules. We use the photon DAs required in the present
calculation up to twist four, following Ref.~\cite{Ball:2002ps}.}

The leading-twist two-particle photon DA is taken in the asymptotic form
\begin{equation}
\phi_\gamma(u)
=
6u\bar u.
\label{eq:app-phi}
\end{equation}
The twist-3 vector DA is
\begin{align}
{\psi^{(V)}(u)}
={}&
-20u\bar u(2u-1)
\notag\\
&+
\frac{15}{16}
\left(
\omega_\gamma^A-3\omega_\gamma^V
\right)
u\bar u(2u-1)
\left[
7(2u-1)^2-3
\right].
\label{eq:app-psiv}
\end{align}
For the twist-4 two-particle DAs, we use
\begin{align}
{\mathcal A(u)}
={}&
{40u^2\bar u^2}
\left(
3\kappa-\kappa^+ +1
\right)
\notag\\
&+
8
\left(
\zeta_2^+-3\zeta_2
\right)
\Big[
u\bar u(2+13u\bar u)
\notag\\
&\hspace{2.0cm}
+
{2u^3(10-15u+6u^2)\ln u}
+
{2\bar u^3}
(10-15\bar u+6\bar u^2)
\ln\bar u
\Big],
\label{eq:app-A}
\\
\mathcal B(u)
={}&
40(1+3\kappa^+)
\int_0^u d\alpha\,
(u-\alpha)
\left[
-\frac12+\frac32(2\alpha-1)^2
\right].
\label{eq:app-B}
\end{align}

The gluonic twist-4 three-particle photon DAs are
\begin{align}
\mathcal S(\alpha_i)
={}&
30\alpha_g^2
\Big[
(\kappa+\kappa^+)(1-\alpha_g)
\notag\\
&\hspace{1.7cm}
+
(\zeta_1+\zeta_1^+)
(1-\alpha_g)(1-2\alpha_g)
\notag\\
&\hspace{1.7cm}
+
\zeta_2
\left\{
3(\alpha_{\bar s}-\alpha_s)^2
-
\alpha_g(1-\alpha_g)
\right\}
\Big],
\label{eq:app-S}
\\
\widetilde{\mathcal S}(\alpha_i)
={}&
-30\alpha_g^2
\Big[
(\kappa-\kappa^+)(1-\alpha_g)
\notag\\
&\hspace{1.7cm}
+
(\zeta_1-\zeta_1^+)
(1-\alpha_g)(1-2\alpha_g)
\notag\\
&\hspace{1.7cm}
+
\zeta_2
\left\{
3(\alpha_{\bar s}-\alpha_s)^2
-
\alpha_g(1-\alpha_g)
\right\}
\Big],
\label{eq:app-Stilde}
\\
\mathcal T_1(\alpha_i)
={}&
-120(3\zeta_2+\zeta_2^+)
(\alpha_{\bar s}-\alpha_s)
\alpha_{\bar s}\alpha_s\alpha_g,
\label{eq:app-T1}
\\
\mathcal T_2(\alpha_i)
={}&
30\alpha_g^2
(\alpha_{\bar s}-\alpha_s)
\Big[
\kappa-\kappa^+
\notag\\
&\hspace{2.3cm}
+
(\zeta_1-\zeta_1^+)(1-2\alpha_g)
+
\zeta_2(3-4\alpha_g)
\Big],
\label{eq:app-T2}
\\
\mathcal T_3(\alpha_i)
={}&
-120(3\zeta_2-\zeta_2^+)
(\alpha_{\bar s}-\alpha_s)
\alpha_{\bar s}\alpha_s\alpha_g,
\label{eq:app-T3}
\\
\mathcal T_4(\alpha_i)
={}&
30\alpha_g^2
(\alpha_{\bar s}-\alpha_s)
\Big[
\kappa+\kappa^+
\notag\\
&\hspace{2.3cm}
+
(\zeta_1+\zeta_1^+)(1-2\alpha_g)
+
\zeta_2(3-4\alpha_g)
\Big].
\label{eq:app-T4}
\end{align}

In addition to these gluonic distributions, the background-field expansion
contains the electromagnetic three-particle photon DAs
$\mathcal S_\gamma$ and $\mathcal T_4^\gamma$. {Their explicit forms are}
\begin{align}
\mathcal S_\gamma(\alpha_i)
&=
60\alpha_g^2
\left(
\alpha_s+\alpha_{\bar s}
\right)
\left[
4-7
\left(
\alpha_s+\alpha_{\bar s}
\right)
\right],
\label{eq:app-Sgamma}
\\
\mathcal T_4^\gamma(\alpha_i)
&=
60\alpha_g^2
\left(
\alpha_{\bar s}-\alpha_s
\right)
\left[
4-7
\left(
\alpha_s+\alpha_{\bar s}
\right)
\right].
\label{eq:app-T4gamma}
\end{align}

{The photon-DA parameters used in the numerical analysis are
summarized in Table~\ref{tab:app-photon-params}.}
\begin{table}[t]
\centering
\caption{{Photon distribution-amplitude parameters at the
renormalization scale $\mu=1~\GeV$. The twist-3 and twist-4 parameters are
adopted from Refs.~\cite{Ball:2002ps,Rohrwild:2007yt}; in particular, the
values $\kappa=0.15$ and $\kappa^+=-0.05$ are taken from
Ref.~\cite{Rohrwild:2007yt}.}}
\label{tab:app-photon-params}
\small
\setlength{\tabcolsep}{2pt}
\renewcommand{\arraystretch}{1.05}
\begin{tabular}{@{}l@{\hspace{8pt}}c@{}}
\toprule
Parameter & Value \\
\midrule
$f_{\gamma,s}^{\perp}$ & $-55.1(1.9)~\MeV$
{\cite{Bacchio:2024pij}} \\
$f_{3\gamma}$ & $-(4.0\pm2.0)\times10^{-3}~\GeV^2$ \\
$\omega_\gamma^A$ & $-2.1\pm1.0$ \\
$\omega_\gamma^V$ & $3.8\pm1.8$ \\
$\kappa$ & $0.15$  \\
$\kappa^+$ & $-0.05$  \\
$\zeta_1$ & $0.40$ \\
$\zeta_1^+$ & $0$ \\
$\zeta_2$ & $0.30$ \\
$\zeta_2^+$ & $0$ \\
\bottomrule
\end{tabular}
\end{table}

\section{\texorpdfstring{{Explicit expressions for the
three-particle invariant amplitudes
$\widehat{\Pi}_{A(B)}^{\rm 3p,\gamma}$ and
$\widehat{\Pi}_{A(B)}^{\rm 3p,g}$}}{Explicit expressions for the
three-particle invariant amplitudes}}
\label{app:basis-amplitudes}


%
For the electromagnetic three-particle photon DAs, we define
\begin{align}
\mathcal F_A^{\gamma}(\alpha_i,v)
&=
(1-2v)\mathcal S_\gamma(\alpha_i)
-
\mathcal T_4^\gamma(\alpha_i),
\notag\\
\mathcal F_B^{\gamma}(\alpha_i)
&=
\frac{m_Q}{m_Q+m_s}
\left[
\mathcal S_\gamma(\alpha_i)
-
\mathcal T_4^\gamma(\alpha_i)
\right].
\label{eq:3p-electromagnetic-weights}
\end{align}

The corresponding three-particle electromagnetic contributions are
\begin{align}
\widehat{\Pi}_A^{\rm 3p,\gamma}
&=
e_Q\ssbar
\left(
e^{-m_Q^2/M^2}
-
e^{-s_0/M^2}
\right)
\left\{
\mathcal J_{\bar{u}_0}
\!\left[
\mathcal F_A^\gamma
\right]
+
\left.
2\frac{d}{dz}
\mathcal L_z
\!\left[
\mathcal T_4^\gamma
\right]
\right|_{z={\bar u}_0}
\right\},
\notag\\
\widehat{\Pi}_B^{\rm 3p,\gamma}
&=
e_Q\ssbar
\left(
e^{-m_Q^2/M^2}
-
e^{-s_0/M^2}
\right)
\frac{m_Q}{m_Q+m_s}
\left\{
\mathcal J_{{\bar u}_0}
\!\left[
\mathcal S_\gamma-\mathcal T_4^\gamma
\right]
+
\left.
2\frac{d}{dz}
\mathcal L_z
\!\left[
\mathcal T_4^\gamma
\right]
\right|_{z={\bar u}_0}
\right\},
\label{eq:pi-3p-photon}
\end{align}
where
\begin{align}
\mathcal J_z[X]
&=
\int_0^z dv
\int_0^{(1-z)/(1-v)}d\alpha_g\,
X\!\left(
1-z-(1-v)\alpha_g,\,
z-v\alpha_g,\,
\alpha_g,\,
v
\right)
\notag\\
&\quad+
\int_z^1 dv
\int_0^{z/v}d\alpha_g\,
X\!\left(
1-z-(1-v)\alpha_g,\,
z-v\alpha_g,\,
\alpha_g,\,
v
\right),
\label{eq:three-particle-J}\\
\mathcal L_z[X]
&=
{\int_0^{1-z}d\alpha_s
\int_0^{\alpha_s}d\alpha'_s
\int_0^z d\alpha_{\bar s}\,
\frac{z-\alpha_{\bar s}}
{(1-\alpha_s-\alpha_{\bar s})^2}
X\!\left(
\alpha_s,\,
\alpha_{\bar s},\,
1-\alpha_s-\alpha_{\bar s}
\right)}
\notag\\
&\quad-
{\int_0^{1-z}d\alpha_s
\int_z^{1-\alpha_s}d\beta\,
\frac{z}{\beta^2}
\int_0^\beta d\alpha'_{\bar s}\,
X\!\left(
\alpha_s,\,
\alpha'_{\bar s},\,
1-\alpha_s-\alpha'_{\bar s}
\right)}
\notag\\
&\quad+
{\int_0^{1-z}d\alpha_s\,
\frac{z}{(1-\alpha_s)^2}
\int_0^{\alpha_s}d\alpha'_s
\int_0^{1-\alpha'_s}d\alpha_{\bar s}\,
X\!\left(
\alpha'_s,\,
\alpha_{\bar s},\,
1-\alpha'_s-\alpha_{\bar s}
\right)}.
\label{eq:three-particle-L}
\end{align}

The gluonic three-particle contributions are determined by the functions
\begin{align}
\mathcal F_{A,\sigma}^{g}(\alpha_i)
&=
\mathcal S
+
\widetilde{\mathcal S}
-
\mathcal T_1
-
\mathcal T_2
+
\mathcal T_3
+
\mathcal T_4,
\notag\\
\mathcal F_A^{g}(\alpha_i,v)
&=
\mathcal F_{A,\sigma}^{g}(\alpha_i)
+
2v
\left(
-\mathcal S-\mathcal T_3+\mathcal T_2
\right),
\notag\\
\mathcal F_B^{g}(\alpha_i)
&=
\frac{m_Q}{m_Q+m_s}
\mathcal F_{A,\sigma}^{g}(\alpha_i).
\label{eq:3p-gluon-weights}
\end{align}
The corresponding invariant amplitudes are
\begingroup

\begin{align}
\widehat{\Pi}_A^{\rm 3p,g}
&=
e_s\ssbar
\left(
e^{-m_Q^2/M^2}
-
e^{-s_0/M^2}
\right)
\mathcal J_{{\bar u}_0}
\!\left[
\mathcal F_A^{g}
\right],
\notag\\
\widehat{\Pi}_B^{\rm 3p,g}
&=
e_s\ssbar
\left(
e^{-m_Q^2/M^2}
-
e^{-s_0/M^2}
\right)
\mathcal J_{{\bar u}_0}
\!\left[
\mathcal F_B^{g}
\right].
\label{eq:pi-3p-gluon}
\end{align}
\endgroup

\section{Explicit expressions for the invariant amplitudes
\texorpdfstring{$\widehat{\Pi}_{AA}$, $\widehat{\Pi}_{AB}$, and
$\widehat{\Pi}_{BB}$}{PiAA, PiAB, and PiBB}}
\label{app:two-point-amplitudes}

In this appendix, we give the OPE expressions used to determine the
decay constants $f_1$ and $f_2$ of the
axial-vector mesons from
Eqs.~\eqref{f1} and \eqref{f2}.

\begin{align}
\widehat{\Pi}_{AA}
={}&
\int_{(m_Q+m_s)^2}^{s_0}
ds\,
e^{-s/M^2}
\rho_{AA}(s)
\notag\\
&+
e^{-m_Q^2/M^2}
\Bigg\{
\langle\bar s s\rangle
\left[
m_Q
-
\frac{m_sm_Q^2}{2M^2}
+
\frac{m_s^2m_Q^3}{2M^4}
\right]
\notag\\
&\qquad
-
\langle\bar s g_s\sigma\!\cdot\!G s\rangle
\frac{m_Q^3}{4M^4}
\Bigg\},
\label{app-piAA}
\\
\widehat{\Pi}_{AB}
={}&
\int_{(m_Q+m_s)^2}^{s_0}
ds\,
e^{-s/M^2}
\rho_{AB}(s)
\notag\\
&+
e^{-m_Q^2/M^2}
\Bigg\{
\frac{\langle\bar s s\rangle}{m_Q+m_s}
\left[
m_Q^2
+
\frac{m_sm_Q}{2}
\left(
1-\frac{m_Q^2}{M^2}
\right)
\right.
\notag\\
&\qquad\qquad\left.
+
\frac{m_s^2}{2}
\left(
\frac{m_Q^4}{M^4}
-
\frac{m_Q^2}{M^2}
-
1
\right)
\right]
\notag\\
&\qquad
-
\frac{
\langle\bar s g_s\sigma\!\cdot\!G s\rangle
}{
4(m_Q+m_s)
}
\left(
\frac{m_Q^4}{M^4}
-
\frac{m_Q^2}{M^2}
-
1
\right)
\Bigg\},
\label{app-piAB}
\\
\widehat{\Pi}_{BB}
={}&
\int_{(m_Q+m_s)^2}^{s_0}
ds\,
e^{-s/M^2}
\rho_{BB}(s)
\notag\\
&+
e^{-m_Q^2/M^2}
\Bigg\{
\frac{\langle\bar s s\rangle}{(m_Q+m_s)^2}
\left[
m_Q^3
+
m_sm_Q^2
\left(
1-\frac{m_Q^2}{2M^2}
\right)
\right.
\notag\\
&\qquad\qquad\left.
+
m_s^2
\left(
\frac{m_Q^5}{2M^4}
-
\frac{m_Q^3}{M^2}
\right)
\right]
\notag\\
&\qquad
-
\frac{
\langle\bar s g_s\sigma\!\cdot\!G s\rangle
}{
(m_Q+m_s)^2
}
\left(
\frac{m_Q^5}{4M^4}
-
\frac{m_Q^3}{2M^2}
\right)
\Bigg\},
\label{app-piBB}
\end{align}
where 

\begin{align}
\rho_{AA}(s)
&=
\frac{\sqrt{\lambda(s, m_Q^2, m_s^2)}}{8\pi^2}
\left[
2
-
\frac{m_Q^2+m_s^2+6m_Qm_s}{s}
-
\frac{(m_Q^2-m_s^2)^2}{s^2}
\right],
\\
\rho_{AB}(s)
=
\rho_{BA}(s)
&=
\frac{\sqrt{\lambda(s, m_Q^2, m_s^2)}}{8\pi^2}
\frac{
3(m_s-m_Q)
\left[
(m_Q+m_s)^2-s
\right]
}{
(m_Q+m_s)s
},
\\
\rho_{BB}(s)
&=
-\frac{\sqrt{\lambda(s, m_Q^2, m_s^2)}}{8\pi^2}
\frac{
\left[
(m_Q+m_s)^2-s
\right]
\left(
2m_s^2-4m_Qm_s+2m_Q^2+s
\right)
}{
(m_Q+m_s)^2s
}.
\end{align}

\bibliographystyle{utcaps_mod}
\bibliography{all}

\end{document}